\documentclass[12pt]{article}
\usepackage{scicite}
\usepackage{times}
\usepackage{graphicx}
\usepackage{multirow}
\usepackage{color}
\usepackage[english]{babel}
\usepackage{amsmath,
amssymb,
amsfonts}
\usepackage{bm}

\newcommand{\no}{{\mathrm{NO}}}
\newcommand{\oo}{{\mathrm{O}_2}}

\title{Observation of correlated excitations in bimolecular collisions}

\author
{Zhi Gao,$^{\ast}$ Tijs Karman,$^{\ast}$ Sjoerd N. Vogels, Matthieu Besemer, Ad van der Avoird,\\ Gerrit C. Groenenboom$^{\ast\ast}$ and Sebastiaan Y.T. van de Meerakker$^{\ast\ast}$\\  \\
\\
\normalsize{Radboud University, Institute for Molecules and Materials}\\
\normalsize{Heijendaalseweg 135, 6525 AJ Nijmegen, the Netherlands}\\
\normalsize{$^{\ast}$ Who contributed equally to this work;}\\
\normalsize{$^{\ast\ast}$ To whom correspondence should be addressed;}\\
\normalsize{E-mail: basvdm@science.ru.nl, gerritg@theochem.ru.nl}}

\begin{document}
\date{\today}

\maketitle

\begin{abstract}
Whereas collisions between atoms and molecules are largely understood, collisions between two molecules have proven much harder to study.
In both experiment and theory, our ability to determine quantum state-resolved bimolecular cross sections lags behind their atom-molecule counterparts by decades. For many bimolecular systems, even rules of thumb -- much less intuitive understanding -- of scattering cross sections are lacking.
Here, we report the measurement of state-to-state differential cross sections on the collision of state-selected and velocity-controlled nitric oxide (NO) radicals and oxygen (O$_2$) molecules.
Using velocity map imaging of the scattered NO radicals,
the full product-pair correlations of rotational excitation that occurs in both collision partners from individual encounters are revealed.
The correlated cross sections show surprisingly good agreement with quantum scattering calculations using \emph{ab initio} NO$-$O$_2$ potential energy surfaces.
The observations show that the well-known energy-gap law that governs atom-molecule collisions does not generally apply to bimolecular excitation processes,
and reveal a propensity rule for the vector correlation of product angular momenta.
\end{abstract}

\newpage

Rotationally inelastic scattering experiments represent a widely-used method to investigate the interactions between molecules.
The crossed molecular beam technique,
in which jets of particles intersect to induce the collisions,
has emerged as the method of choice to probe the interactions with the highest possible level of detail.
Precise control of the initial conditions combined with accurate determination of the post-collision properties permit a particularly sensitive probe of these interactions.

For atom-molecule systems, a plethora of experimental methods is available to precisely control the molecule's initial quantum state \cite{Kramer:JCP42:767}, velocity \cite{Meerakker:CR112:4828,Meerakker:NatPhys4:595}, as well as its orientation \cite{Stolte:NAT353:391}, and to probe the relevant properties after the collision.
State-to-state cross sections \cite{Chandler:book}, steric effects \cite{Aoiz:PCCP2015,Wang:NatChem4:636}, as well as vector correlations between pre and post-collision properties \cite{Costen:CSR37:732,Brouard:PRL111:183202,Brouard:JCP135:084305} can be precisely measured,
and in general show excellent agreement with predictions based on full quantum mechanical calculations.
During the last decades, this wealth of experimental and theoretical studies has revealed how energy and angular momentum is transferred between the collision partners, although surprising discoveries continue to be made \cite{Onvlee:NatChem9:226}. Sophisticated qualitative and intuitive models have been developed that explain general trends in the scattering cross sections, which can also be determined quantitatively by near-exact quantum mechanical treatments.

By contrast, the study of bimolecular collisions at the full quantum state-resolved level is virtually uncharted territory,
and only little is known about excitation processes that can occur when two molecules collide.
As opposed to an atomic target,
a molecular collision partner possesses internal degrees of freedom, which
yields the possibility that excitations are induced in both collision partners in a single encounter.
One of the most intriguing aspects of bimolecular collisions relates to the so-called product pair correlations of these excitations, i.e., for a collision event between molecule $A$ and $B$ in which molecule $A$ is excited to a given final state,
what is the probability that molecule $B$ is also excited to specific final states?
These, as well as other basic questions relating to the amount of energy transferred to each collision partner and the role of monomer rotational angular momenta are at present unanswered. For instance, it is unclear if rotational excitation is most likely to occur in both molecules by equal amounts, or how the sense of rotation of the collision products are coupled to each other.

Rotational product pair correlations contain a wealth of new information on collision mechanisms and the underlying interactions,
and their experimental study will yield unprecedented opportunities to extend our precise knowledge of atom-molecule collisions to more complex bimolecular interactions, both on a qualitative intuitive level and on a quantitative computational level. Yet, formidable challenges exist to study energy transfer and excitation correlations in bimolecular systems\cite{Chefdeville2012},
in particular for open shell species such as NO and O$_2$ \cite{Kirste:Sience338:1060}. On the theoretical side, the interaction is described by multiple potential energy surfaces (PESs) with nonadiabatic couplings between them.
Furthermore, the steep scaling of the computational effort of quantum scattering calculations impedes numerically exact treatment of bimolecular scattering -- which is feasible for atom-molecule systems -- especially for open-shell species with additional electronic and spin degrees of freedom.
Experimentally, it is challenging to prepare both molecular reagents with a sufficient density and state purity to observe excitation in both collision partners.
In addition, even if these conditions are met, it is difficult to probe the correlations between the excitation cross sections.
Simply measuring the product states of both species individually only yields total state-to-state cross sections,
but does not reveal information on the correlations \cite{liu:07}.

Elegant methods in principle exist to measure these correlated excitations without explicitly detecting both fragments.
The idea is that if one can accurately measure the internal state and kinetic energy of collision product $A$,
one can infer the simultaneous excitation of an \emph{unobserved} collision partner $B$ from a change in product $A$ kinetic energy (see Fig. 1).
In the early 1980's,
rotational product pairs have been probed for HD$-$D$_2$ collisions \cite{buck:83} using time-of-flight spectroscopy,
where detection of the correlated channels was enabled by the relatively large rotational energy splitting of the D$_2$ molecule.
Using the powerful velocity map imaging (VMI) technique,
vibrational product pairs have been studied recently for photodissociation processes \cite{Townsend:Science306:1158,Hause:Natchem3:932,Grubb:science335:1075,Zhang:JCPA118:2413} and in reactive scattering \cite{Lin:Science300:966,Yan:Science316:1723,Zhang:Science325:303,Wang:Science331:900,Wang:Science342:1499,Yang:Science347:60}.
With VMI,
state-selective detection of one product species yields a set of concentric rings that reveal the pair-correlated cross sections,
offering a kinematically complete picture of the scattering process \cite{liu:07} (see also Fig. 1).
Insight into the reaction mechanisms is obtained from the various scattering channels that are often fully resolved in the images due to the relatively large energetic spacing of vibrational modes.
For rotationally inelastic scattering, however,
the much smaller energy spacings between rotational states of typically only a few cm$^{-1}$ imply that the rings are very close together. Resolving the individual rings requires exceptional experimental resolutions, which are often not available in crossed beam experiments due to the velocity and angular spreads of the reagent beams.
Recently, several state of the art experiments were performed to study collisions of NO and ammonia with several molecular collision partners \cite{brouard:17,luxford:17,gijsbertsen:06,yang:11,tkac:14,tkac:15},
however, the direct experimental observation of rotational product pairs has remained elusive.

Here, we report the measurement of fully resolved rotational product-pairs for inelastic collisions between NO ($X\,^2\Pi$) radicals and O$_2$ ($X\,^3\Sigma_g^-$) molecules in a crossed beam experiment.
We combined Stark deceleration and velocity map imaging to record state-resolved integral and differential cross sections with an exceptional resolution \cite{Zastrow2014,Vogels2015},
which allowed us to fully resolve multiple concentric rings in the scattering images pertaining to the correlated rotational excitation in both molecules.
Trends in the intensities of product-pair excitations reveal rules of thumb for bimolecular excitations in this system,
akin to the well-known energy-gap law that applies to atom-molecule collisions.
The observations are understood in terms of short-ranged head-on and long-ranged glancing collision mechanisms,
and a propensity rule for the vector correlation of fragment angular momenta is discovered.
The pair correlated integral and differential cross sections show excellent agreement with the cross sections from quantum mechanical coupled-channels scattering calculations based on high-level \emph{ab initio} PESs.

We chose collisions between NO and O$_2$ radicals as a model system for
several reasons. The NO radical, in its $^2\Pi$ electronic ground state,
has been a benchmark system for rotational energy transfer studies for
decades. The rotational constant of the O$_2$ molecule in its $^3\Sigma$
ground state is similar to that of NO, such that rotational excitation
in one species is likely to be accompanied by rotational excitation in
the other. Rotationally inelastic scattering between NO and O$_2$ has
been studied in several crossed beam experiments in which NO product
state and angular distributions were measured
\cite{Bacon:JCP108:3127,brouard:17}, aiding our search for pair
correlations. Furthermore, the energetic spacing between rotational
levels is sufficiently small that many correlated excitations may be
expected, and sufficiently large that all scattering channels can be
resolved using the Stark decelerator.

\section*{Results}

The rotational energy level diagrams of both NO and O$_2$ are shown in Fig. 1.
The rotational levels of NO and O$_2$ are labeled by the quantum number $j_\no$ and $N_\oo$, respectively, that are further split into a nearly-degenerate $\Lambda$-doublet (for NO, labeled $e$ and $f$) and closely spaced fine-structure components (for O$_2$, labeled $j_\oo$).
A Stark decelerator was used to prepare monochromatic packets of state-selected NO [$X\,^2\Pi_{1/2}, v=0, j_\no=1/2, f$, referred to hereafter as ($1/2f$) \cite{Note:1}] radicals with a velocity of 450 m/s and negligible velocity spread. The labels $X^2\Pi_{1/2}$ and $v$ indicate the electronic and vibrational state of the NO radical, respectively.
A beam of neat O$_2$ molecules with a velocity of 700 m/s and a velocity spread of 50~m/s (FWHM) was produced using a room temperature Even-Lavie valve.
The state purity of the reagent beams was probed using laser ionization detection schemes.
Before the collision, $>99\%$ of the NO radicals reside in the ($1/2f$) level,
and $> 90$\% of the O$_2$ molecules reside in the $N_\oo=1$ rotational ground state,
of which 23\%,18\% and 59\% is found in the $j_\oo=0, 1$, and $2$ levels, respectively.

Inelastic scattering between NO and O$_2$ was studied at a collision energy of 160 cm$^{-1}$ and an energy spread around 20 cm$^{-1}$ (FWHM),
using a crossed beam geometry with a 45$^{\circ}$ crossing angle.
State-selective detection in combination with VMI was used to probe the scattered NO radicals.
Scattering images (shown in Fig. 2) were measured for excitation of NO into eight final states,
ranging from $j^{\prime}_\no = 1/2$ to $13/2$. For each final state probed,
a set of four concentric rings was observed corresponding to different excitations of the O$_2$ collision partner to various levels $N^{\prime}_\oo$.
Excitations to individual fine structure components ($j'_\oo)$ could not be resolved,
but the signature of $j'_\oo$-resolved cross sections was apparent in the widths and intensities of the different rings (\emph{vide infra}).
The outer ring with the largest diameter corresponded to scattering events in which the NO radical was excited but the O$_2$ molecule scattered elastically and remained in the $N_\oo=1$ state.
The inner rings with decreasing diameters corresponded to excitations of O$_2$ into the $N^{\prime}_\oo$ = 3, 5, and 7 levels.

The correlations between ($j^{\prime}_\no,N^{\prime}_\oo$) scattering cross sections were investigated further from the radial scattering distributions within a narrow window of scattering angles near forward scattering,
where the images showed largest intensity [see Supplementary Information: Analysis and Simulation] (Fig.~3).
In this figure, the radii of the spheres corresponding to excitation of O$_2$ into the possible $N^{\prime}_{\oo}$ as predicted from the kinematics of the experiment are indicated.
A strong correlation between excitation of NO and O$_2$ was observed.
For the ($1/2f$) to ($1/2e$) transition in NO, which is energetically elastic as no translational energy is transferred to rotation of the NO radical,
the majority of the scattering flux was found in the outer ring with largest diameter,
indicating that the majority of the O$_2$ molecules also scattered elastically.
This is consistent with the well-known energy-gap law of atom-molecule scattering,
which states that cross sections decrease with the amount of energy transferred.
For higher excitations of NO, the intensities of the inner rings progressively increase and were found to become comparable to or even exceed the intensity of the outer ring.
The correlated excitation cross sections observed here for $\no - \oo$ seem to violate the energy gap law;
strong excitation in both the $\no$ and $\oo$ molecules is preferred over excitation in one collision partner only, although the total amount of energy transfer is higher.

\section*{Discussion}

To corroborate these findings, we performed coupled-channels scattering calculations of the DCSs for product-pair excitations of $\no-\oo$.
We calculated \emph{ab initio} interaction potentials which were diabatized \cite{karman:16a} to account for non-adiabatic interactions in the $\no-\oo$ system, consisting of two open-shell fragments. The $\oo$ fine-structure was treated using a recoupling approach, as described in the Supplementary section "Theory". The theoretical pair-correlated DCSs were used to simulate images that were compared directly to the experimental results, see Fig.~2 and Fig.~3. These simulations took the kinematics of the experiment into account, as well as all fine-structure resolved $\no (j_\no=1/2f) + \oo (N_\oo=1, j_\oo) \rightarrow \no (j^{\prime}_\no) + \oo (N^{\prime}_\oo, j^{\prime}_\oo)$ transitions that could contribute to the experimental images. Both the angular distributions and the relative intensities of each ring in the experimental and simulated images show excellent agreement.

We further analyzed the theoretical results to gain insight into the observed scattering dynamics.
From the calculated opacity functions (see Supplementary Information: Theory) we could interpret the rigorous quantum scattering calculations in terms of classical impact parameters,
such that one can quantify the contribution of short and long-ranged collisions.
This analysis showed that the inelastic channels where either or both $\no$ and $\oo$ are excited to higher rotational states are governed by short-ranged head-on collisions.
Cross sections for nearly elastic channels,
with low excitation energies for both $\no$ and $\oo$,
are dominated by long-ranged glancing collisions.
Only this long-ranged contribution follows the energy gap law as its intensity decreases with the energy transferred to rotation of both $\no$ and $\oo$.
This explains the violation of the energy gap law that is observed for highly-excited $\no$ channels.
If a highly excited $\no$ state is probed, the dynamics is short-ranged irrespective of the $\oo$ state such that the product pair excitations have comparable cross sections and are readily observed.
If a nearly-elastic $\no$ channel is probed, the $N_\oo=1$ elastic channel has an additional long-ranged contribution and hence a much larger cross section,
such that inelasticity in $\oo$ only plays a minor role.

In order to understand what happens in these bimolecular
collisions in more detail, we also investigated the relative direction of the pre and
post collision rotational angular momenta. We introduced $\vec{j}_{AB} =
\vec{j}_{\no} + \vec{N}_{\oo}$ and $\vec{j'}_{AB} = \vec{j'}_{\no} +
\vec{N'}_{\oo}$ as the vector sum of the monomer angular momenta before
and after the collision, respectively, and calculated partial cross
sections restricted to specific values for the corresponding quantum
numbers $j_{AB}$ and $j'_{AB}$. For the initial states with $j_\no=1/2$
and $N_\oo=1$, the possible values for $j_{AB}$ are $1/2$ and $3/2$. The
$j_{AB} \rightarrow j'_{AB}$ partial cross sections showed a strong
propensity towards the highest value of $j'_{AB}$ permitted by the
triangular inequality (see Fig. 4 for the $j'_\no=11/2e$
state as an example), indicating that the monomer angular momenta are
stretched such that both rotationally excited molecules exhibit the same
sense of rotation after collision. This vector correlation can be
understood classically, as equal and opposite forces acting on the atoms
that approach most closely exert equal torques on both molecules (see
Fig. 4), independent of whether the forces are attractive
or repulsive. This result does not imply orientation nor alignment of
the individual molecules, but constitutes a vector correlation between
the product molecular angular momenta. We also investigated whether the
two colliding molecules exhibit not only classical correlations between
the excited states, but form a quantum entangled pair that persists
after the collision. We found that the angular momenta of two excited
molecules indeed become quantum entangled, leading to a violation of
Bell's inequalities that may in principle be probed experimentally ({\it
see Supplementary section: Entanglement}). This entanglement occurs especially
for the Fraunhofer diffraction oscillations in the differential cross
sections at small scattering angles \cite{Onvlee:NatChem9:226}.

Our experiments show that energy transfer processes for bimolecular collisions can now be studied with resolutions necessary to probe correlated excitation processes. As a first foray into this field, our joint experimental and theoretical study on rotational energy transfer between state-selected NO and O$_2$ molecules has proven the ideal model system to unravel how excitations between two colliding molecules are correlated, both in terms of the amount of energy transferred, and in terms of vector correlations of the angular momenta. The success attained here implies that experiments on a wide variety of bimolecular systems can now be performed. This yields unprecedented opportunities to study scattering processes not occurring in atom-molecule collisions, such as resonant energy transfer processes between energetically near-degenerate levels, or the stereodynamics of mutually oriented systems. Ultimately, the ability to monitor bimolecular collisions with resolutions as demonstrated here allows us to extend our near-perfect understanding of atom-molecule interactions to bimolecular systems.

\section*{Methods}

The experiments were performed using a crossed molecular beam apparatus which has been described previously \cite{Onvlee2014,Vogels2015}.
A mixture of 5\% $\no$ seeded in Kr at a typical pressure of 1 bar is expanded through a Nijmegen Pulsed Valve \cite{Yan2013} and subsequently enters a Stark decelerator.
The $\no$($X\,^2\Pi_{1/2}, j=1/2, \emph{f}$) package was selected by the Stark decelerator with a speed of 450 m/s.
After exiting the decelerator and 69 mm free flight, the package of $\no$ collides with another package of $\oo$ molecules at 45$^\circ$ angle of incidence. A (1+1$^{\prime}$) resonance enhanced multiphoton ionization (REMPI) scheme was used to state-selectively detect the reagent and scattered $\no$ radicals using two pulsed dye laser systems.
The Velocity Map Imaging (VMI) optics is used to collect the $\no$ ions after the ionization.
The ion optics used in the experiment consists of three extractor plates as used by Suits and coworkers \cite{Townsend2003}.
The grounded time of flight (TOF) tube has a length of 1100 mm, and
the voltage applied on repeller and extractor plates are 3000V, 2755V, 2519V and 2160V respectively.
The collision energy was calibrated using the method described in Refs.~\cite{Vogels2015,Onvlee2014,Zastrow2014}.
The pixel-to-m/s conversion factor of the VMI setup was calculated by using the extremely well calibrated velocities of $\no$ radicals which emerge from the Stark decelerator.
From the radii of the scattering images, we can determine the mean collision energy to be 160 cm$^{-1}$.

The $\oo$ beam was produced by expanding pure $\oo$ with a backing pressure of 5 bar into vacuum using a room temperature Even-Lavie valve \cite{Even2014}.
The initial rotational state's distribution of the $\oo$ pulsed beam was probed using a (2+1) REMPI scheme, from which we conclude that the $\oo$ molecules emerge from the Even-Lavie valve predominantly in the rotational state $N_\oo=1$. See Supplementary section "Experimental Methods" for more details.

\textbf{Data availability.} The data that support the findings of this study are available from the corresponding author upon reasonable request.

\section*{Acknowledgements}
The research leading to these results has received funding from the European Research Council
under the European Union's Seventh Framework Programme (FP7/2007-2013) / ERC grant
agreement 335646 MOLBIL. This work is part of the research program of the Netherlands
Organization for Scientific Research (NWO). The expert technical support by Niek Janssen, Andr\'e van Roij, and Edwin Sweers is gratefully acknowledged.\\

\textbf{Author contributions}\\
The experiments were conceived by S.Y.T.v.d.M.. The experiments were carried out by Z.G. with the help of S.N.V.. Data analysis and simulations were performed by Z.G.. Potential energy surfaces were calculated by T.K., A.v.d.A., and G.C.G.. Scattering calculations were performed by T.K. and M.B.. Analysis of quantum entanglement was performed by T.K.. All authors were involved in the interpretation of the data, discussed the results, and commented on the manuscript. The paper was written by Z.G., T.K. and S.Y.T.v.d.M. with contributions from all authors.\\

\begin{figure}
\begin{center}
\includegraphics[width=\textwidth]{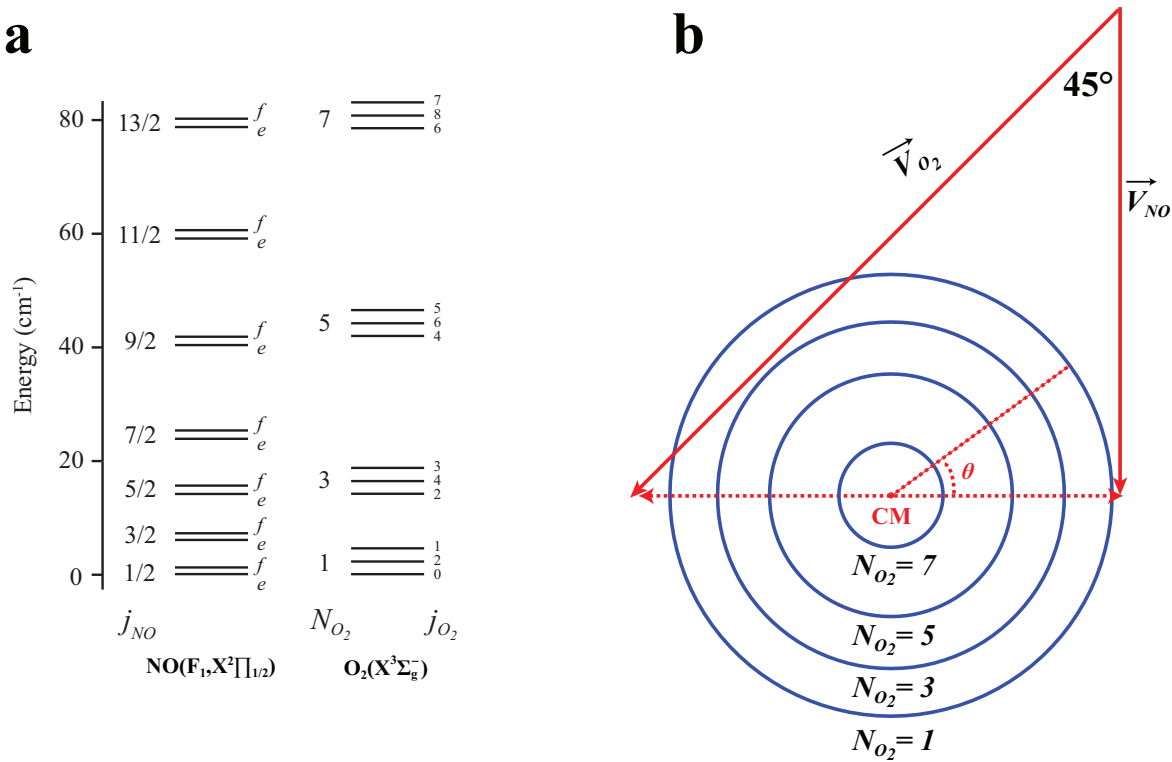}
\caption{
{\bf Rotational energy level diagrams of NO and O$_2$}, and the Newton diagram for NO-O$_2$ collisions.
Each rotational level in NO and O$_2$ is labelled in the energy diagram (a) by the quantum number $j_{\no}$ and $N_{\oo}$,
respectively.
Due to nuclear spin statistics of the two identical O atoms,
only levels with an odd value for $N_{\oo}$ exist.
The $\Lambda$-doublet splitting of each rotational level in NO and the spin-orbit splitting of each rotational level in O$_2$ is greatly exaggerated for reasons of clarity.
(b) Schematic representation of the appearance of rotational product pairs in rotationally inelastic collisions between NO radicals and O$_2$ molecules.
Before the collision,
the NO and O$_2$ molecules reside in the rotational ground state $j_{\no}=1/2,f$ and $N_{\oo}=1$,
respectively,
and approach each other with velocities $\vec{v}_{\no}$ and $\vec{v}_{\oo}$ in the laboratory frame.
Upon collision,
the NO and O$_2$ molecules can get rotationally excited to rotational levels $j^{\prime}_{\no}$ and $N^{\prime}_{\oo}$.
In the center of mass (CM) frame,
scattered NO radicals lie on a sphere with a radius determined by the combined rotational energy that is taken up by both the NO and O$_2$ molecules.
Different combinations of ($j^{\prime}_{\no},
N^{\prime}_{\oo}$) excitations result in a set of nested spheres,
whose projections onto a two dimensional plane result in a series of concentric rings when probed using VMI.
}
\end{center}
\end{figure}

\begin{figure}
\begin{center}
\includegraphics[width=\textwidth]{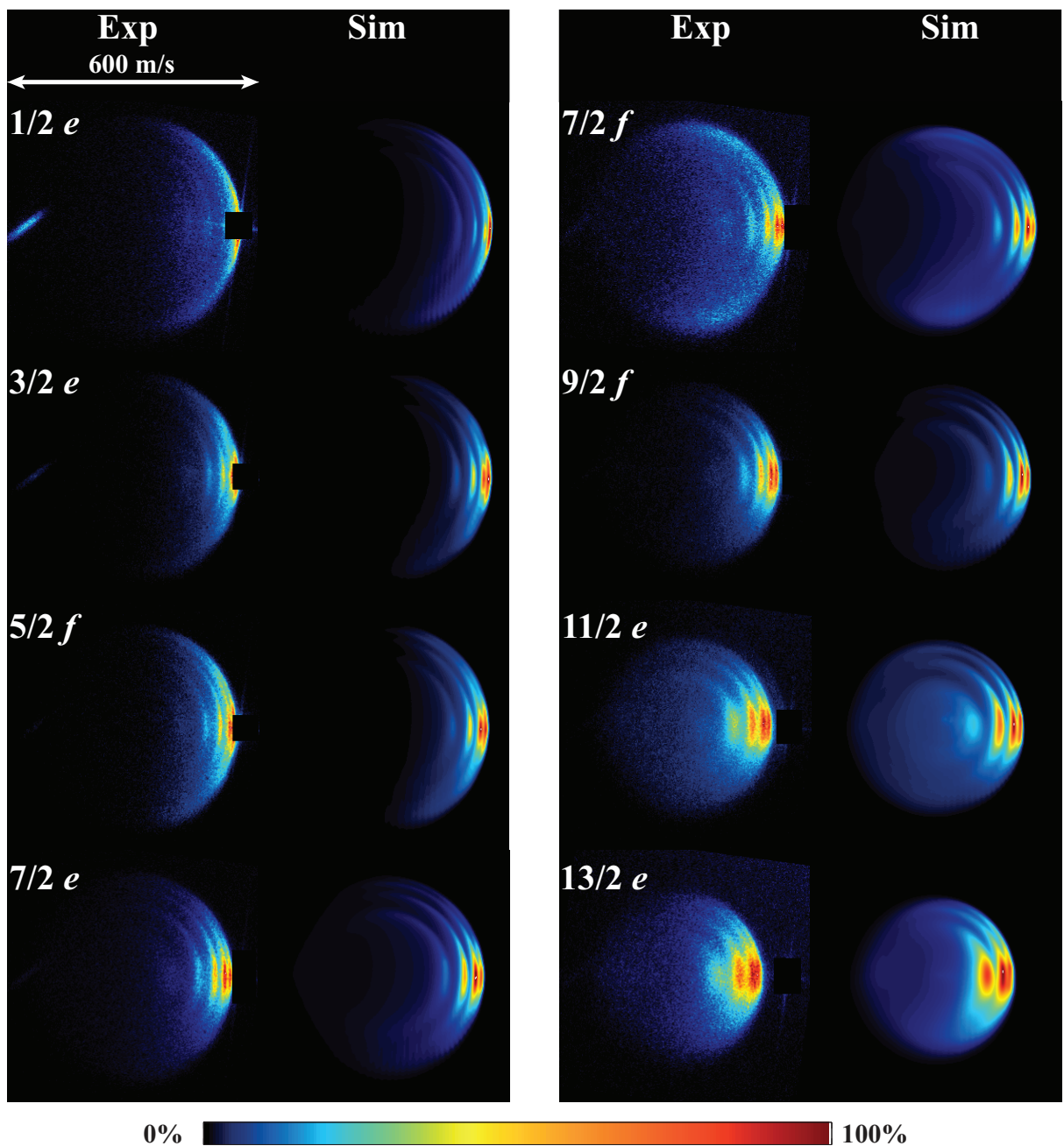}
\caption{
{\bf Experimental (\textbf{Exp}) and simulated (\textbf{Sim}) scattering images} for the scattering processes NO ($1/2f$) + O$_2$ ($N_{\oo}=1$) $\rightarrow$ NO ($j^{\prime}_{\no}$) + O$_2$ (N$^{\prime}_{\oo}$).
Images are presented such that the relative velocity vector is directed horizontally,
with forward scattered angles positioned at the right side of the image.
Small segments of the images around forward direction are masked because of the imperfect state selection of NO packet. The images display multiple concentric rings pertaining to correlated excitations in both the NO radicals and O$_2$ molecules.
}
\end{center}
\end{figure}

\begin{figure}
\begin{center}
\includegraphics[width=\textwidth]{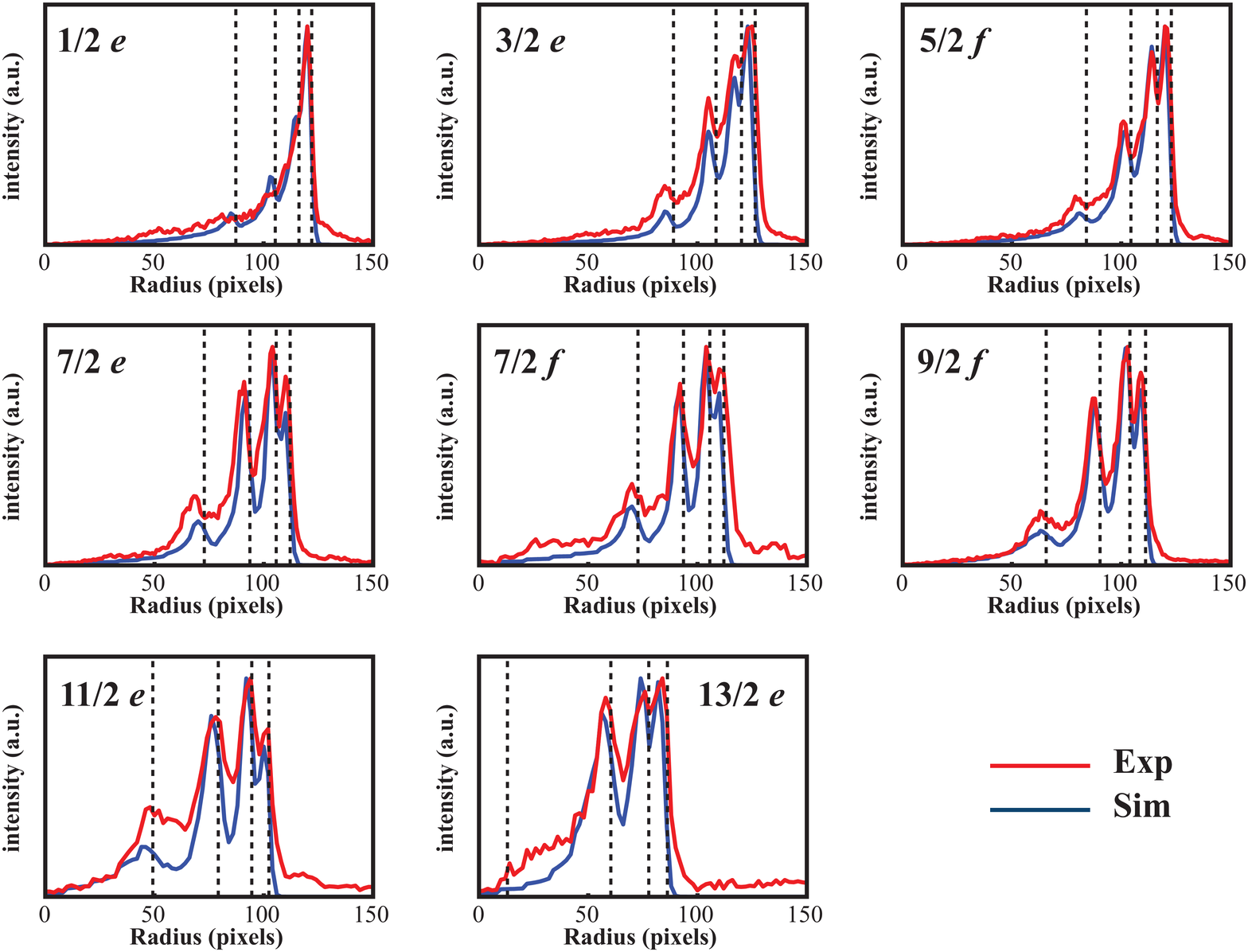}
\caption{
{\bf Radial intensity distribution} of the experimental (red curves) and simulated (blue curves) scattering images from Figure 2. The kinematic cutoff radii for various product-pairs are indicated by vertical dashed lines. For higher excitations of NO, the intensities of the inner rings progressively increase, indicating that strong excitation in NO is accompanied by strong excitation in O$_2$.
}
\end{center}
\end{figure}

\begin{figure}
\begin{center}
\includegraphics[width=0.495\textwidth]{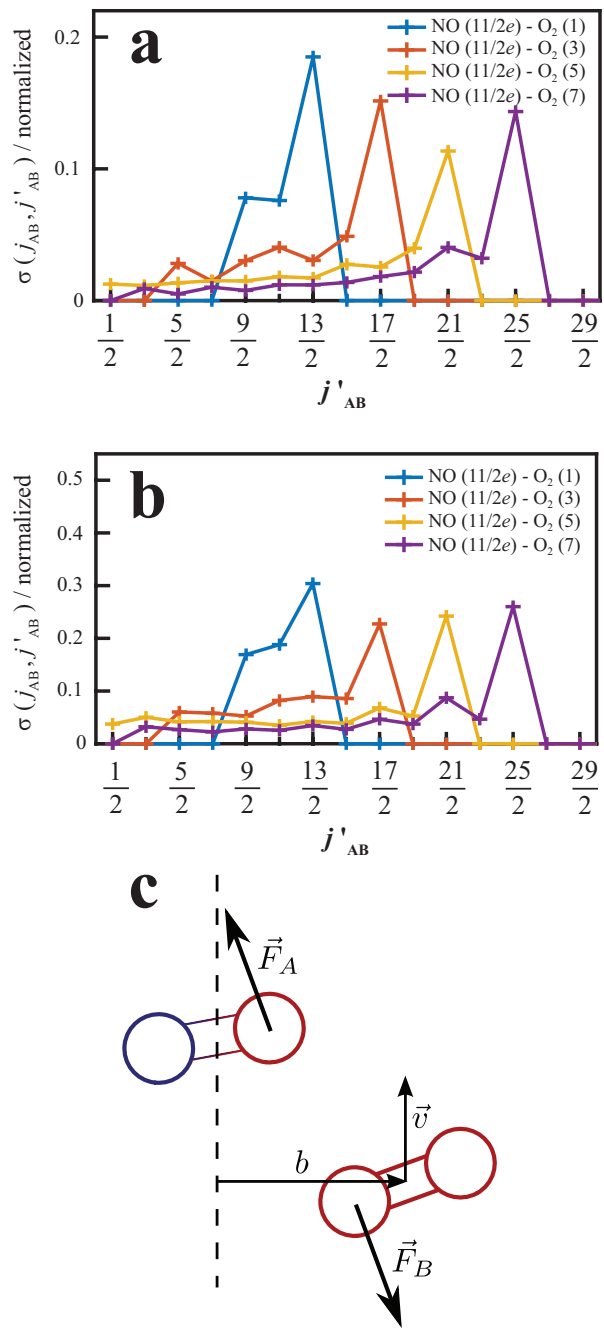}
\caption{
{\bf Analysis of the sum of angular momenta} before and after the collision. (a) Partial cross sections $j_{AB} = 1/2 \rightarrow j^{\prime}_{AB}$ and (b) $j_{AB} = 3/2 \rightarrow j^{\prime}_{AB}$ pertaining to the scattering process NO ($1/2f$) + O$_2$ ($N_{\oo}=1$) $\rightarrow$ NO ($j^{\prime}_{\no} = 11/2e$) + O$_2$ (N$^{\prime}_{\oo}$), showing a propensity towards the highest value of $j^{\prime}_{AB}$ permitted. This indicates that both the NO and O$_2$ molecules exhibit the same sense of rotation after the collision. (c) Schematic of two diatomic molecules that approach each other with relative velocity $\vec{v}$ and impact parameter $b$ in a coplanar geometry, and that interact with repulsive forces $\vec{F_A}$ and $\vec{F_B}$.
}
\end{center}
\end{figure}


\end{document}